# Vibrating Cantilever Transducer Incorporated in Dual Diaphragms Structure for Sensing Differential Pneumatic Pressure

Akin Cellatoglu[1] and Balasubramanian Karuppanan[2]

[1]Department of Comp. Engineering, European University of Lefke, North Cyprus, Turkey
acellatoglu@eul.edu.tr
[2]Department of EE Engineering, European University of Lefke, North Cyprus, Turkey
kbala@eul.edu.tr

## ABSTRACT

*Pneumatic pressure cells with thin metallic spherical diaphragm of shallow spherical shell configuration linked with vibrating wire pickup or vibrating cantilever pickup were reported in the past. In order to enhance the sensitivity of the pressure cell this work considers dual diaphragm structure fitted with cantilever pickup. The design and development of the pressure cell with this dual diaphragm structure having cantilever pickup is presented here. The geometrical design is optimally made as to sense either mono pressure or differential pressure resources. The cantilevers of the two diaphragms are excited to produce vibrations and the frequencies of vibrations are determined by picking up signals from orthogonally arranged opto-coupler links. With the computed frequency a lookup table is referred to obtain the pressure acting on the concerned diaphragm. In the external circuits, the average pressure and the differential pressure acting on two diaphragms are computed. Furthermore transmitting circuits taking the average pressure and differential pressure in digital form and analogue form to remote area are presented. Performance analysis of the proposed mechatronic pressure cell is made and its improved performance over other pressure cells is presented.*

## KEYWORDS

*Cantilever excitation, diaphragm with cantilever, differential pressure transducer, dual diaphragm cell, pressure transmitter*

## 1. INTRODUCTION

Measurement of mono pneumatic pressure and differential pneumatic pressure are widely being practised in production plants and process industries [1-3]. These types of pressure measuring devices invariably use diaphragm based structures with attachment of appropriate pickups [4]. While vibrating wire pickup is popular for displacement measurement [5], pressure cells with circular metallic diaphragms fitted with vibrating wire pickup were also reported in the past [6,7]. Improving the sensitivity of diaphragm pressure cells to a greater extent and improving further its dynamic performances, a shallow spherical shell diaphragm fitted with either optical pickup, inductive pickup or capacitive pickup was reported recently [8]. These types of pressure cells have specific operating pressure ranges depending upon the size and metallic properties of the diaphragms employed. To enhance the sensitivity with vibrating wire pickup a dual diaphragm structure with a vibrating wire stretched at the vertices was also attempted [9]. A diaphragm installed pressure cell attached with a cantilever excitation for electromechanical conversion was also reported [10]. This has more durability compared to vibrating wire pickup device. Presently, as to enhance further the performance of the diaphragm based pressure cell with cantilever

DOI : 10.5121/ijsc.2011.2409　　　　95

pickup, we have developed cantilever installed dual shallow spherical diaphragms and its structure and performance are presented. The design for optimized geometrical structure for fixing the dual pickups occupying the minimal space is described. The dynamic performances are analyzed and their improvement over single diaphragm transducer is presented.

### 1.1. Cantilever Attachment with Single Diaphragm

A relatively thick circular metallic diaphragm having a thin metallic strip formulating as cantilever brazed at its center was used to sense mono pneumatic pressure [10]. The strip was loosely hinged at a pivoted support at a small distance from the diaphragm such that when the vertex of the diaphragm drifts due to applied pressure, the strip could smoothly slide through the hinge. The vibrations were produced by electromagnet excitation on the strip and an optocoupler at orthogonal direction was used to sense the signal produced due to vibration. As the frequency of vibration would be dependent on the length of the strip extending beyond the hinge it represents the pressure acting on the diaphragm. By measuring the frequency the pressure is computed and calibrated.

### 1.1.1. Frequency Pressure Relationship

When the pressure is applied to the diaphragm it makes the strip fixed at the centre to move. If the strip of length $l$ extending beyond the pivoted hinge point is subjected to vibrations then it begins to oscillate. By using proper distances between the vertex and hinge and the metallic properties of the strip the oscillations are limited to first order oscillations as to concede clarity in the signal picked up the opto-coupler. The magnitude of oscillations $y(x)$ seen at different distances $x$ was obtained as

$$y(x) = \left\{\cosh\frac{1.875}{l}x - \cos\frac{1.875}{l}x - 0.734\left\langle\sinh\frac{1.875}{l}x\right\rangle\right\}\frac{1}{2}y_1\cos\left\langle\frac{3.515}{l^2}\sqrt{\frac{EI}{\gamma}}t\right\rangle \quad (1)$$

where $E$: Modulus of rigidity, $I$: Moment of inertia and $\gamma$ : mass of the strip per unit length. The factor $k$ is a constant depending on the length, mass and other mechanical properties. Therefore, the vertex displacement depends on the pressure acting on the diaphragm which in turn would affect the frequency of vibration of the metallic strip. Ultimately, by measuring the frequency from the opto-coupler we can calibrate it to pressure acting on the diaphragm.

## 2. USE OF SHALLOW SPHERICAL SHELL DIAPHRAGM FOR PRESSURE SENSING

According to thin plate theory, a thin circular metallic diaphragm having elastic properties undergoes changes in its geometrical surface when force is applied on it [11-13]. This change in deformation depends on the geometrical size and boundary conditions. The deflection would be the maximum at the centre. With the use of a shallow spherical shell the displacement at the vertex is proved be more compared to flat diaphragm [8]. As enhanced drifts ensure increase in sensitivity the structure yielding larger drift is always desirable for developing the pressure cell.

### 2.1. Drift and Pressure Relations of Shallow Spherical Shell

A shallow spherical shell used for sensing mono pressure is shown in Figure 1 in simplified form. The shallow shell has the radius $ra$ and height $f$. The height at vertex is the maximum and zero at rim. At any radial distance $r$ the height of the shell observed is denoted by $w$.

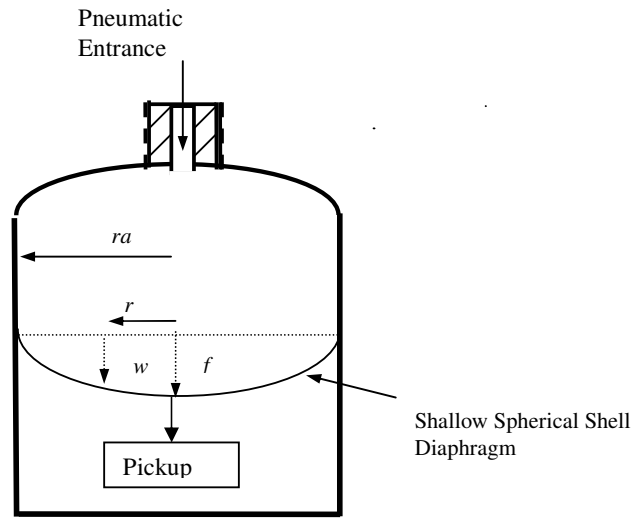

Figure 1. Schematic of diaphragm based pressure cell

Depending upon the pressure acting on the diaphragm and deformation happening on the surface the magnitude of $w$ and $f$ would change [8]. The following relations show the dependancy of f and w on the applied pressure.

$$f = A - \frac{ra}{3.A} \quad (2)$$

where

$$A = \left[ \frac{\eta}{2} + \left( \frac{\alpha^3}{27} + \frac{\eta^2}{4} \right)^{1/2} \right]^{1/3} \quad (3)$$

and

$$\alpha = \frac{56.h^2}{(1+\gamma)(23-9.\gamma)} \quad (4)$$

where h : plate thickness and $\gamma$ : Poisson ratio.

$$\eta = \frac{7.p.ra^4 h^2}{8.D(1+\gamma)(23-9.\gamma)} \quad (5)$$

where D is flexural rigidity given by

$$D = \frac{E.h^3}{12.(1-\gamma^2)} \quad (6)$$

$p$ is pressure acting on diaphragm and $E$ : Young's modulus

The altitude of the diaphragm $w$ at any radial distance $r$ is related to $f$ as

$$w(r) = f.\left[ 1 - \left( \frac{r}{ra} \right)^2 \right]^2 \quad (7)$$

When the pressure acting on the diaphragm changes, these heights $f$ and $w$ would change as per the above relations.

With the appropriate pickup linked to the vertex the drift in vertex can be transformed into an electrical signal representing the pressure acting on the diaphragm.

## 3. DUAL DIAPHRAGM STRUCTURE FOR PRESSURE CELL

With the use of two diaphragms the dual diaphragm structure helps enhancing the sensitivity since each diaphragm moves due to application of pressure contributing the displacement nearly two times of single diaphragm cell. In this way, dual diaphragm structure helps in sensing the differential pressure also.

### 3.1. Dual Diaphragm Structure with Vibrating Wire Pickup

The dual diaphragm structure is now described with a vibrating wire incorporated in it. Figure 2 shows the schematic of the pressure cell having two diaphragms and a wire bonding their vertices. The diaphragms are fixed rigidly inside a metallic hollow cylinder at an estimated spacing. The wire is set into vibration by mechanical force exerted magnetically. The frequency of the vibration is measured by an opto-coupler in which the light projected into the photo sensor is interrupted by the vibrating wire. The frequency is then calibrated for the pressure acting on the diaphragms. The directions of the magnetic vibration exciter and the opto-coupler around the vibrating wire are kept mutually orthogonal to each other.

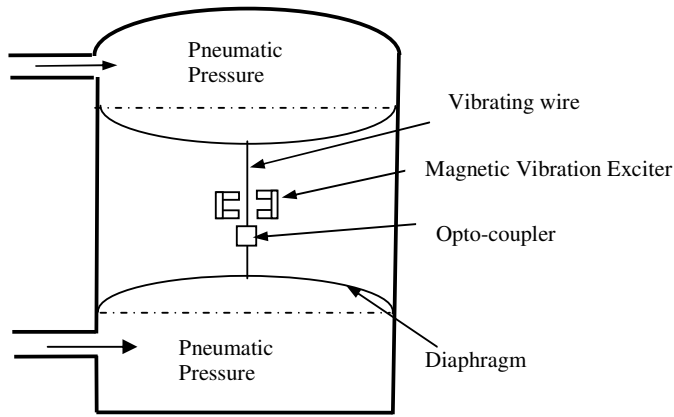

Figure 2. Schematic of dual diaphragm implanted vibrating wire pressure transducer

### 3.1.1. Frequency Pressure Relations

The frequency of the vibrating wire stretched on the diaphragms depends on the pressure acting on the diaphragms. A tightly stretched string of length $L$ at the extreme ends is subject to set into vibration by small force acting on it. The forces experienced on the string in orthogonal directions are given below.

$$Fy = T.Sin(\theta + \Delta\theta) - T.Sin\theta \qquad (8)$$

$$Fx = T.Cos(\theta + \Delta\theta) - T.Cos\theta \qquad (9)$$

where $T$ is the tension involved.

The partial differential equations concerning the transverse motions in terms of space and time are as follows.

$$\frac{\partial^2 y}{\partial x^2} = \frac{\mu}{T}\frac{\partial^2 y}{\partial t^2} \tag{10}$$

*where*

$$\mu = \frac{m}{L} \tag{11}$$

where *m* is mass of the wire in Kg and L is length in meters.

The solution of the partial differential equation can be shown as

$$r(x) = K.Sin\left\{\omega\left(\sqrt{\frac{\mu}{T}}\right)x\right\} \tag{12}$$

Applying boundary conditions at *x=0 and x=L* we can show that

the fundamental frequency of oscillations as

$$f = \frac{1}{2L}\sqrt{\frac{T}{\mu}} \tag{13}$$

With the use of two diaphragms experiencing vector added tensions *T1 and T2* the frequency f becomes

$$f = \frac{1}{2L}\sqrt{\frac{(T1+T2)q}{\mu}} \tag{14}$$

where *q* is the alignment factor having its value slightly less than 1.

If A is the area of the diaphragm then tension acting on the diaphragm is

$$T = p.A \tag{15}$$

where p is the pressure acting on the diaphragm.

With *T1 and T2* remaining the same, the frequency *f* of the string stretched between two diaphragms becomes

$$f = \frac{1}{2L}\sqrt{\frac{2\ pAq}{\mu}} \tag{16}$$

Therefore, for a given applied pressure to the two tracks of the pressure chamber, the frequency of the vibrating wire stretched between two diaphragms is √2 times that of the single diaphragm based pressure cell. This enhancement in *f* contributes to similar enhancement in sensitivity. On the other hand, with this structure we can not sense the differential pressure since single frequency is sensed from single wire. When two pressure sources are applied to the two tracks of the cell, the resulting frequency picked up would be proportional to the average of the two pressure sources.

### 3.2. Dual Diaphragm Cell with Cantilever Pickup

The vibrating wire pickup attachment has a specified range of pressure measurement whereas cantilever pickup has its operating range larger than that of vibrating wire cell. Although the dynamic performance of the vibrating wire is relatively good its durability is limited due to fatigue materialized in the vibrating wire. Cantilever operated pickup is mechanically stable and rugged yielding indefinite life time. Moreover, we can sense differential pressure since each diaphragm is fitted with a cantilever and the vibration in two cantilevers are sensed to compute the differential pressure.

#### 3.2.1. Geometrical Structure

The geometrical structure of dual diaphragm cell with cantilever is optimally designed as to accommodate the dual magnets and dual opto-couplers for producing excitations of cantilevers and interference free optical readout. Figure 3 shows the geometrical schematic of the pressure cell sensing the differential pressure. There are two hollow cylinders holding the metallic diaphragms joined to form a cell. In the vertices of the diaphragms thin metallic strips working as cantilevers are bonded. In order to avoid overlapping and mechanically interrupting each other the strips of the vertices have to be laterally displaced. Therefore, the two hollow cylinders holding the diaphragms are laterally shifted and brazed at the rims making the joint. A circular disc contacting the rim of left cylinder (Disc1) holds the opto-coupler kept orthogonal to the strip. This would pickup a signal proportional to vibration. A similar disc is fixed at the rim of right cylinder (Disc2) also for picking up the vibration of the strip. Figure 4 gives the end view of the discs showing the location of opto-coupler and opening for the strip of the other.

As seen in Figure 3 there are two electromagnets employed for exciting the vibrations in the strips. Both strips are excited simultaneously by driving a pulse to the coils of electromagnets which would attract the strips and release them producing the vibrations. Connections to coils in the electromagnets and to opto-couplers are made through fine fibre like wires taken along the surfaces and extended to the connectors.

The two sources of pressures denoted as *p1* and *p2* are entering into two pneumatic tracks of the pressure cell. Source *p2* enters in the central track (Track 2) and impede on the diaphragm 2 resulting in drift of its vertex. Source *p1* enters into the outer track (Track 1) surrounding the cylinder holding diaphragms and impede on the diaphragm 1. This causes drift in its vertex in opposite direction of the drift of diaphragm 2. The drift in the vertex and hence the frequency of vibration of the strips depend on the pressure acting on the concerned diaphragm.

#### 3.2.2. Relationships of Various Parameters

The geometrical sizes of the diaphragms and the housing contributing to overall size of the pressure have a definite operating range of the pressure. The diaphragm used in the proto type unit is a medium strength Al alloy of 2.5cm radius, 1.0mm thickness and 0.5cm of height for the spherical shell with Young's modulus of $210 GN/m^2$ and Poisson ratio as 0.3. This has a pressure range of 100 to 200 Pascal. The strip used is of thickness 0.5mm and length 4cm. The distance between the vertex of the diaphragm and the pivot under maximum pressure conditions is 0.5cm.

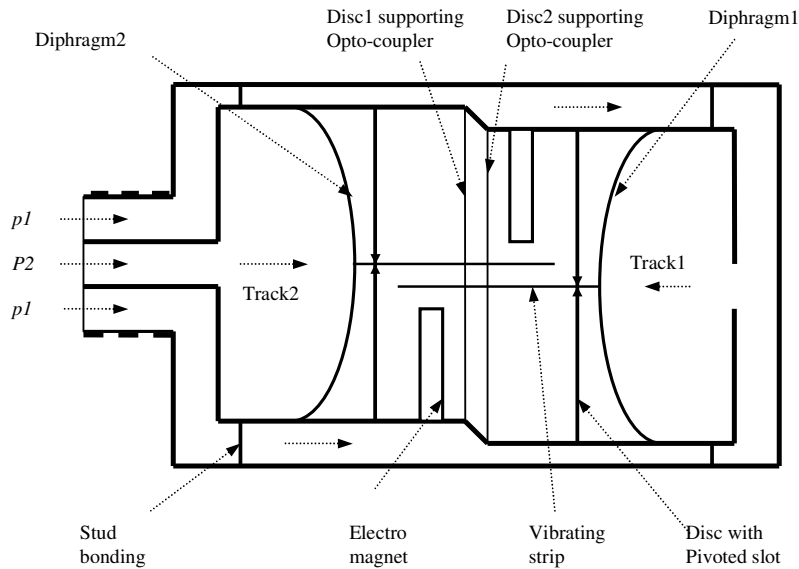

Figure 3. Structure of the differential pressure sensor

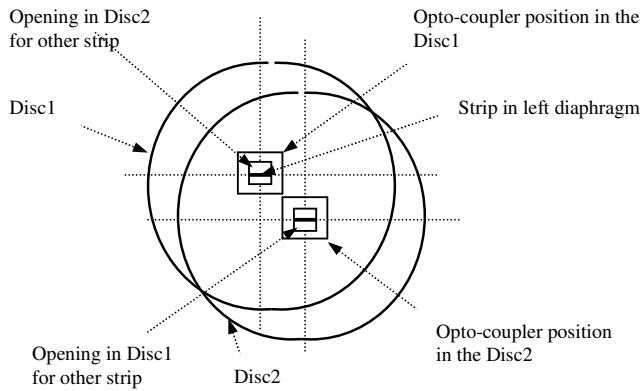

Figure 4. End view showing the position of opto-couplers and passage for the strip of other diaphragm

Table 1 gives samples of information showing the relationship between the pressures, drift in vertex fixing the length of the strip and the frequency of vibration. Figure 5 shows the graphical diagram illustrating pictorially the relationships between the applied pressure and the length of the strip extending the pivot. The length of the strip would be responsible for the frequency of vibration and Figure 6 shows the relationship between the applied pressure and the frequency of vibration. These relationships would vary when the geometrical conditions or materials are changed.

International Journal on Soft Computing ( IJSC ) Vol.2, No.4, November 2011

Table 1. Pressure, length and frequency relations

| Pressure (Pascals) | Length extending beyond the pivot (Cm) | Frequency ( Hz) |
| --- | --- | --- |
| 100 | 2.50 | 250 |
| 120 | 2.65 | 236 |
| 140 | 2.85 | 222 |
| 160 | 3.04 | 210 |
| 180 | 3.25 | 200 |
| 200 | 3.50 | 191 |

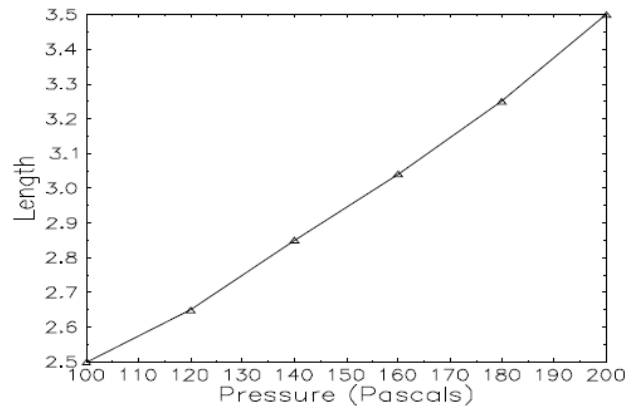

Figure 5. Pressure vs length of cantilever

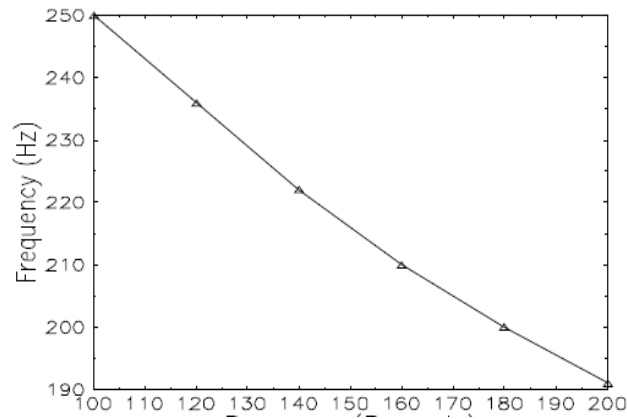

Figure 6. Pressure vs frequency of vibration

### 3.2.3 Instrumentation Setup for Sensing Pressure and Processing

Intel's 8086 family of microprocessors are being extensively used in general purpose computer systems such as PCs and as single board microcontrollers. A microcontroller established with microprocessor 80286 is the used in this project for driving the signals for exciting electromagnets and reading the sensed signals from opto-couplers and processing them.    Figure 7 shows the simplified schematic and circuit diagram of the instrumentation unit. The Clock frequency set is 20MHz.

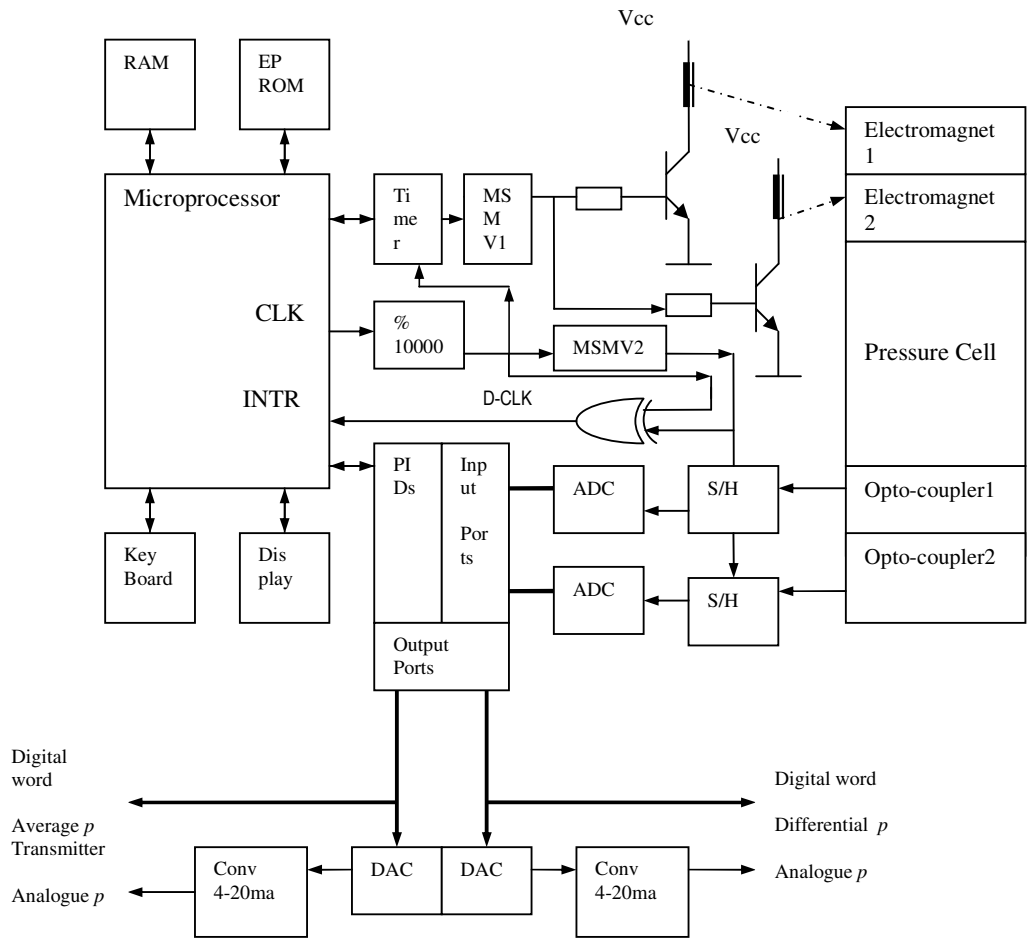

Figure 7. Simplified schematic of the instrumentation scheme for sensing and transmitting pneumatic pressure

The microprocessor is interfaced with PIDs (Peripheral Interface Devices) having programmable I/O ports, EPROM, RAM, Timer, Keyboard and display devices as in standard format and procedures. The timer is programmed to produce a square wave signal of 1Hz which in turn is converted into a signal of 0.4s ON and 0.6s OFF periods by using a monostable multivibrator (MSMV1). The 0.4s ON signal drives a transistor to excite the electromagnet which in turn would attract the solenoid to hit the strip. The same signal drives both electromagnets simultaneously for setting the strips into vibrations. Both the opto-couplers produce signals as per the amplitude and frequency of vibrations. In every cycle of observation the vibration is sensed for the period of 0.6s. In the following 0.4s period there are no vibrations since this time is reserved for the solenoid to hit the strip. In the 0.6s period the decaying sine wave is sampled periodically. The sampling signal is derived from 20MHz clock signal of the microprocessor. Its frequency is externally divided by a factor of 10000 and 2 KHz signal is used as sampling signal (D-CLK). This sampling signal is applied to both the S/H (sample and Hold) circuits extended to opto-couplers. MSMV2 generates mono pulse for driving the S/H circuits. The flash ADCs (Analogue to Digital Converter) used here are built on advanced architecture [14, 15] with high speed and reduced complexity and their outputs are extended to input ports available in the PIDs. The sampled data of opto-coupler is read into microprocessor and processed further.

During the period of sampling and holding the flash ADC would certainly complete its conversion and is ready for feeding to microprocessor. Therefore, the input and output of MSMV2 are given to an Ex-OR circuit to derive the hardware interrupt signal (INTR) for the microprocessor. This would request the microprocessor to read the sample and process further.

From the sampled data gathered from the opto-couplers, the frequencies of vibration are estimated for both the cantilevers and in each case the lookup table is referred to determine the pressure acting on each diaphragm. After then the average pressure and the differential pressure are computed. They are driven to their respective output ports externally. Therefore we have the average and differential pressure in binary form. They are converted into analogue form by driving to DACs (Digital to Analogue Converter). DAC would produce the analogue output in current form and is ready for transmission to a remote location. Electric current transmission avoids losses in the cables taken to remote location. In order to meet the industry standards the electric current level is kept in the range of 4-20mA and standard circuits are used for this purpose.

### 3.2.4. Timing Diagram

The timing diagram would illustrate about the sequence of performed by the microcontroller in the pressure cell. Figure 8 shows essential timings. Figure 8.A shows the time slots for the electromagnet drive and for reading the opto-coupler samples to processor. Figure 8.B shows the time slots for the S/H derived for the frequency divided clock D-CLK in each sampling instant. It also shows the slot for interrupt signal to the microprocessor.

### 3.2.5. Software Overview of the Instrumentation System

The sampled data are gathered and saved in RAM memory on interrupt basis and from which the instantaneous pressure is computed. During the OFF period of 0.6s of electromagnet excitation the sampled data are gathered. With the sampling frequency of 2 KHz, there are 1200 samples available in each cycle. During the ON period of 0.4s where there is no vibration the computation of pressure is carried on. Also display routine is called on every second to display the computed information.

#### 3.2.5.1. Features of Main program

In the main program the following operations are organised. *i.* Computations of frequency of vibration and pressure data from the 1200 reserved locations of RAM memory where sampled data of opto-coupler are available; *ii.* Since two sets of samples are gathered from two opto-couplers there are two sets of data available for computing pressure acting on each diaphragm. After computing the dual pressures ( *p1, p2*), the average pressure $\{(p1+p2)/2\}$ is computed. The average pressure information would be useful in applications where mono source pressure sensing is desired. Also, the differential pressure (*p1-p2*) is computed. This is the main objective of this project to have both the average pressure and the differential pressure.

 The average pressure is useful in measuring mono pressure sources. On the other side differential pressure measurement has its own applications in instrumentation systems. The computed differential pressure and the average pressure are driven to respective output ports. *iii.* A display routine is called during the 0.6s period where the average pressure and differential pressure are displayed alternately.

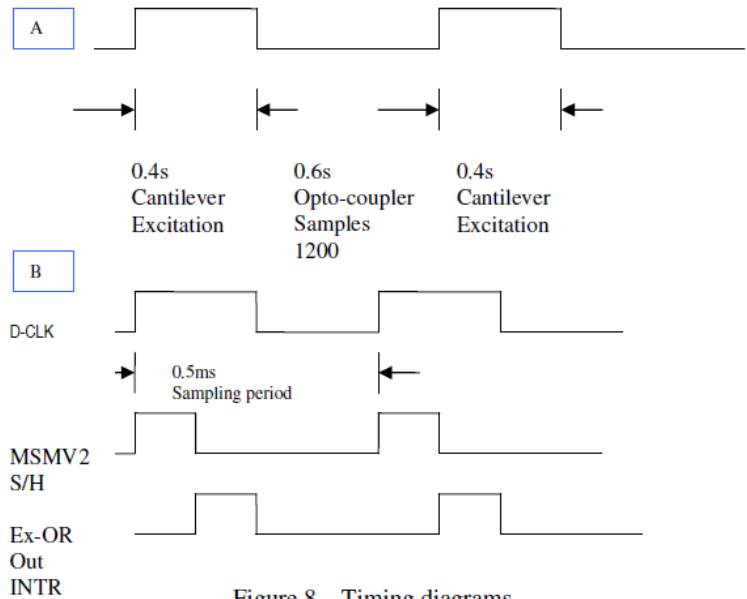

Figure 8. Timing diagrams
A. Time slots of 1Hz signal from MSMV1
B. Opto-coupler Sampling time slots

Computation of the frequency of vibration of each strip is performed by using the 1200 discrete samples and calling an FFT algorithm. The fundamental frequency of oscillation is taken into consideration for using it as an input to lookup table. The lookup table relating the frequency of oscillation to the pressure acting on the diaphragm is made available in the EPROM provided in the microcontroller. For the given diaphragm and the cantilever the lookup table is programmed in the EPROM. Therefore, once the frequency is determined, the pressure acting on the diaphragm is quickly recovered by referring the lookup table.

### 3.2.5.2 Interrupt Service Procedure

The interrupt occurs once in 0.5ms to gather the data from ADC to memory. When interrupt occurs, the binary word from the ADC is read by the microprocessor and saved in the reserved memory locations. Therefore, in the interrupt service procedure the following sequence of operations are performed. *i.* Read the sample from two ADCs one by one and save in the addresses pointed for each. *ii.* Increment the address pointers. If the upper limit of 1200 locations is reached then set the pointer to the starting address *iii.* Return to the main program.

## 4. EXPERIMENTATION

In order to assess the static and dynamic performance of the dual diaphragm pressure cell with cantilever pickup simulated experiment and practical experiment are performed. Simulated experiment is conducted by setting a known pressure and computing various parameters involved in equations (2) to (7) by a computer program. The drift in vertex is computed and for the change in length the frequency of vibration is estimated by using the FFT algorithm. After computing the frequency the pressure acting on the diaphragm is obtained by using the lookup table information available in the EPROM. This computation provides the analytical results.

For conducting the practical experiment the microcontroller used in Figure 7 is extended with an I-to-P converter and the hardware scheme with this extension is shown in Figure 9. With a program written for experimentation the set pressure data is sent to the output port. This is converted into analogue current by a DAC (MC1408) [16] connected at the output port.The I-to-P

105

International Journal on Soft Computing ( IJSC ) Vol.2, No.4, November 2011

Converter converts this into equivalent pneumatic pressure and gets applied to the pressure cell. The average pressure output given by the pressure cell in binary word is taken through an input port to the microcontroller and taken for display by calling the display routine. This makes us to get the response of the pressure cell for the applied pressure.

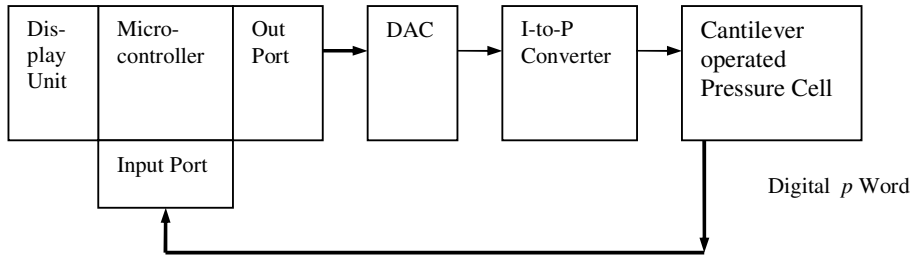

Figure 9. Hardware extension for experimentation

Figure 10 shows the results of analytical computation and practical experimentation. The analytical computation is made only once. This is because the computed pressure has to be the same as applied pressure used in various equations. Once the frequency is determined both the analytical computation and practical experimentation use the same lookup table loaded in the EPROM for computing the pressure. The lookup table has been prepared carefully by involving all geometrical and elastic characteristics of the diaphragm and also the geometrical and mechanical properties of the strip working as cantilever. Therefore, for the given frequency determined analytically or practically we get the same pressure. Nevertheless due to environmental conditions such as temperature the yield from diaphragm might vary in determining the frequency of vibration. Therefore in the case of practical experimentation five different trials at different times have been done and the median of the pressure is taken as the pressure sensed and plotted in Figure 10.

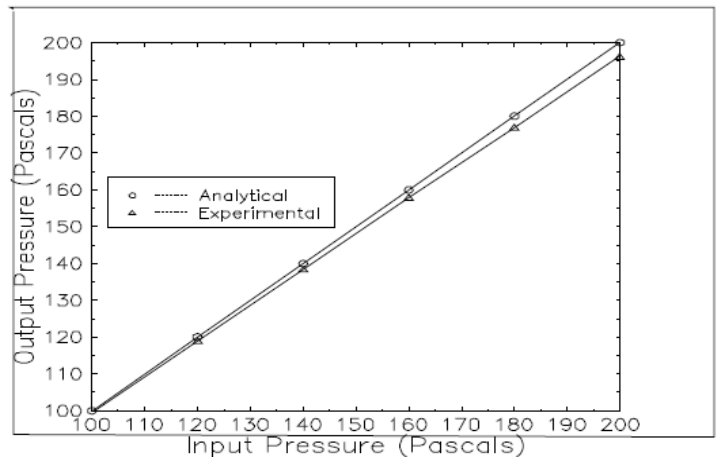

Figure 10. Experimental results for static Input

### 4.1. Testing Dynamic Performance

As the excitation of cantilever and readout procedure from the opto-coupler involves in a timing period of 0.4s followed by 0.6s (Figure 8), after applying the pressure we need to wait at least 0.4s for determining the frequency and the applied pressure. Therefore, a three step staircase pressure signal with increasing magnitudes is given as input and responses are obtained from single diaphragm-cantilever cell and dual diaphragm-cantilever cell. The responses of these pressure cells are plotted as given in Figure 11. The magnitude of the three steps are at 0.5 Pascal, 1 Pascal and 4 Pascal respectively. For the step of 0.5 Pascal, the single diaphragm cell failed to produce significant output as the change in length produced is so low that the change in frequency produced for this step is not good enough for sensing and processing the output.

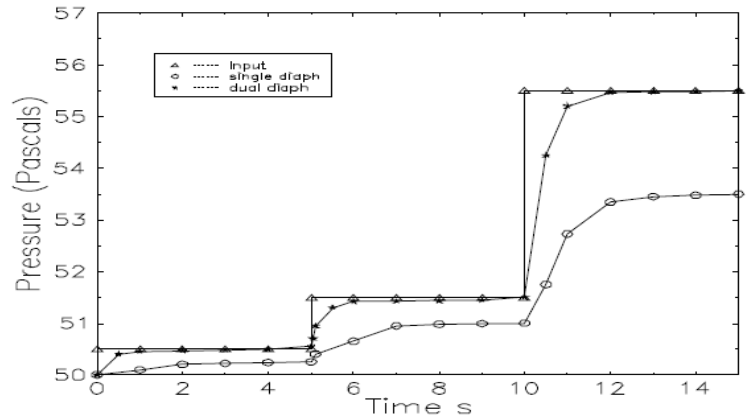

Figure 11. Dynamic performances of pressure cell with cantilever pickups

For dual diaphragm-cantilever cell the addition of two meagre changes in frequencies produce a noticeable output and could be readout without an error. If differential pressure of similar step magnitudes is applied to dual diaphragm cell, the response is found to be similar to that of mono pressure input. As discussed, for mono pressure input the dual pressure cell produces the added version of the signals of two cantilevers. For differential pressure source the magnitude differences of signals from two cantilevers are produced. On the other side, the single diaphragm-cantilever cell has no provisions for sensing the differential pressure as there is only single track of pressure input available. Therefore, the dynamic performance of the dual diaphragm-cantilever cell is much improved compared to single diaphragm-cantilever cell.

## 5. DISCUSSIONS AND CONCLUSION

The dual diaphragm pressure cell has twin metallic diaphragms of identical geometry and materials used for construction. For mono pneumatic pressure measurement with the dual diaphragm-cantilever cell we need to apply the same pneumatic medium as input to both the pneumatic tracks in the chamber. Both pneumatic inputs acting on two diaphragms would produce same strain effects on the diaphragms leading to similar excitations on the vibrating cantilevers. The frequencies of vibrations of both the cantilevers would be almost identical and produce identical pressure output derived from the frequency of vibration. We therefore take the average of the pressure output resulted from two cantilevers as the pressure output of the pressure cell. The sensitivity is enhanced due to the contributions to output from two cantilevers. The differential pressure sensing is possible only with the dual diaphragm-cantilever cell since the single diaphragm-cantilever cell has only one pressure track.

The vibrating wire transducer fixed on the diaphragm also measures frequency and relates it to the pressure acting on the diaphragm. However, vibrating wire transducer is not durable owing to the weak fatigue characteristics of the wire leading to its mechanical breakdown for collapsing. The procedure for replacing the wire with another one involves tedious tasks in alignment procedures involved. Therefore, being rugged cantilever based pickup is preferred in many instances. The dual diaphragm structure holding cantilever or wire enhances the sensitivity. Nevertheless, the differential pressure sensing is feasible with dual diaphragm-cantilever cell only since the vibrating wire transducer has one wire and one optical readout producing only the average pressure output.

The pressure transmitter for the proposed dual diaphragm-cantilever cell is designed to transmit the average pressure and differential pressure computed by microprocessor. The output is available in both digital form and analogue form. This enables us to make the on spot digital display of the sensed pressure and also allows us to take the pressure signal to any distance such as to the control room. Moreover, these signals can easily be fed to any PC for further processing and analysis.

The diaphragm used in the cantilever operated pressure sensor is relatively thicker compared to the ones used in vibrating wire pressure sensor since this structure has to support relatively larger mass acting as physical load to the diaphragm than the thin vibrating wire. Nevertheless, for analytical dealing arriving at the vertex drift comes under the same thin plate theory with governing equations used for the vibrating wire pressure sensor. The lookup table in EPROM used for both the simulated experiment and the practical experiment need be changed whenever there is a change in geometry or change in the properties of materials used for making the diaphragm or the cantilever strip. With enhanced sensitivity, durability and capability to pickup differential pressure the dual diaphragm-cantilever pressure cell finds wide ranges of applications in process industries.

## ACKNOWLEDGEMENTS

The authors would like to thank Rector of European University of Lefke for providing financial support for this project.

**Authors**

Akin Cellatoglu received his Bachelor's degree in Electronics and Communication Engineering from Eastern Mediterranean University, Tu rkish Republic of Northern Cyprus in 1996. He obtained his M.Sc degree and Ph.D degree from the University of Surrey in 1998 and 2003 respectively. Dr.Akin is with the Computer Engineering department of European University of Lefke, Turkish Republic of Northern Cyprus since September 2003. His fields of specialization are in video codec systems, multimedia and communication networks.

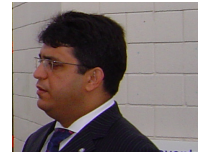

Balasubramanian Karuppanan received his Bachelor's degree in Electronics and Communication Engineering from PSG College of Technology, Madras University, India in 1971. He obtained his M.Tech and Ph.D degrees from the Indian Institute Technology, Madras in 1976 and 1984 respectively. He was working in Calicut University, India, during the periods 19 72-1990 and 1995-1998 in various positions as Lecturer, Asst. Professor and Professor. In 1988, he did post doctoral research at Tennessee Technological University, Cookeville, TN, USA under Fulbright Indo-American Fellowship program. He joined Cukurova University, Adana by June 1990 as Professor and worked there until Feb1995. By 1996, he was granted the Best College Teacher's Award from the University of Calicut, India. From Feb1998 onwards, he is with the Faculty of Architecture and Engineering of European University of Lefke, Turkish Republic of Northern Cyprus. Dr.Balasubramanian is a life member of Instrument Society of India and the Indian Society for Technical Education. His fields of specialization are in 3D imaging and microprocessor based systems.

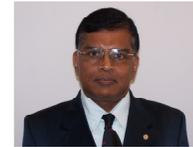